\documentclass{aa}
\usepackage{epsfig}
\def\ltsima{$\; \buildrel < \over \sim \;$}
\def\simlt{\lower.5ex\hbox{\ltsima}}
\def\gtsima{$\; \buildrel > \over \sim \;$}
\def\simgt{\lower.5ex\hbox{\gtsima}}
\def\src{1E{\thinspace}1615+061}
\def\approxgt{\mathrel{\hbox{\rlap{\lower.55ex \hbox {$\sim$}}
        \kern-.3em \raise.4ex \hbox{$>$}}}}
\def\approxlt{\mathrel{\hbox{\rlap{\lower.55ex \hbox {$\sim$}}
        \kern-.3em \raise.4ex \hbox{$<$}}}}

\begin{document}
\thesaurus{4(13.25.2; 
	     11.19.1; 
	     11.09.1 1E1615+061; 
	     02.01.2)} 

\title{Evidence for ionized reprocessing in the X-ray spectrum of the 
Seyfert 1 Galaxy \src}

\author{M. Guainazzi\inst{1,2} \and L. Piro\inst{3} \and 
M. Capalbi\inst{2} \and A.N. Parmar\inst{1} \and 
M. Yamauchi\inst{4} \and M. Matsuoka\inst{5}}

\institute{
{Astrophysics Division, Space Science Department of ESA, ESTEC, Postbus 299,
NL-2200 AG Noordwijk, The Netherlands}
\and
{{\it Beppo-SAX} Science Data Center, c/o Nuova Telespazio, Via Corcolle 19, 
I-00131 Roma, Italy}
\and
{Istituto di Astrofisica Spaziale, C.N.R., Via Fosso del Cavaliere, I-00133 
Roma, Italy}
\and
{X-ray Astronomy Laboratory, Dept. of Electrical and Electronic Engineering, 
Faculty of Engineering, Miyazaki University, 1-1 Gakuen-Kibanadai-Nishi, 
Miyazaki, Miyazaki 889-2155, Japan}
\and
{Cosmic Radiation Laboratory, RIKEN, 2-1 Hirosawa, 350-01 Wako-shi, 
Saitama, Japan} 
}
   
\offprints{M. Guainazzi [mguainaz@astro.estec.esa.nl]}

\date{Received 6 April 1998 ; accepted 28 August 1998}

\maketitle

\markboth{M.Guainazzi et al.}{Ionized reprocessing in the 
X-ray spectrum of \src}

\begin{abstract}

The Seyfert~1 galaxy \src\ was observed to 
display a very steep (${\rm \Gamma \simeq 4.2}$) and intense soft X-ray 
spectrum during a HEAO-1 A2 observation in the 1978.
Such an exceptionally soft X-ray state has never been observed
subsequently, but the source has continued to exhibit a large
(up to a factor 6) range of X-ray intensity variability.
The overall UV/X-ray
spectrum of this source, observed during a multiwavelength campaign 
in 1991--1992, can be well fit with a self-consistent, low-$\dot{\rm M}$ 
accretion disk model. In this model, 
the soft X-rays result were suggested
to arise from reflection of the nuclear emission by
mildly ionized (${\rm \xi \simeq 100}$) material in the
inner regions of the disk. In this {\it Paper} 
we report the results of an ASCA observation in 1995 August,
which give a direct confirmation of such a scenario.
The spectrum may be modeled as a power-law with a photon index
of 1.8, together with absorption consistent with the galactic
line of sight value, substantial reflection from ionized 
material and an iron fluorescent K$_{\alpha}$ emission
line. The centroid energy ($\simeq$6.6--6.8~keV)
implies a ionization stage $\ge$~Fe{\sc xix}. The line profile 
is consistent with that expected from a kinematically and 
gravitationally distorted line around a black hole.
These results provide the first direct evidence for the
existence of considerable amount of ionized material around the
nucleus of a ``broad'' Seyfert~1 galaxy.

\end{abstract}


\section{Introduction}

In many Seyfert 1 galaxies and radio-quiet quasars, 
the extrapolation of the intermediate (2--10~keV) X-ray 
emission to the soft (0.1--2~keV) energy range reveals 
evidence for the presence of soft excesses.
These soft excesses are generally well correlated with 
the optical/UV emission bumps. This supported the hypothesis that
they are the same spectral component
(Walter \& Fink 1993; Puchnarewicz et al. 1996; Laor
et al. 1997), probably thermal emission from an accretion
disk (Czerny \& Elvis 1987).
Attempts to fit disk models to the optical/UV spectra
of single objects have been generally successful
(Laor 1990; Bechtold et al.,
1994; Kuhn et al. 1995), but have been unable to
significantly constrain the emission process due to the large
number of free parameters.
On the other hand, multiwavelength and/or sample studies generally fail
to reproduce the energy spectrum
predicted by disk models (Ulrich \& Molendi 1995;
Laor et al. 1997).
Moreover, the good correlation between
the optical and UV variations observed from 
some Seyfert galaxies (Peterson et al.
1991; Clavel et al. 1991)
argues against the UV originating from 
thermal emission in an accretion disk.

An alternative possibility is to explain the soft and hard X-ray
emission in a common framework. The key concept is
{\it reprocessing}. There is good observational evidence that the
primary continuum emitted in the nuclear regions of active galaxies
is strongly reprocessed by optically thick matter surrounding a
supermassive black hole. The strongest evidence comes 
from the discovery of a ``Compton bump''
and fluorescent K$_{\alpha}$ 
emission features from neutral or mildly ionized iron in the spectra of 
a number of Seyfert~1 galaxies (Pounds et al. 1990;
Piro et al. 1990;
Nandra \& Pounds 1994). The discovery of the gravitationally and
kinematically distorted profile of the iron line in a deep ASCA observation
of the Seyfert~1 galaxy MCG-6-30-15 (Tanaka et al. 1995),
and the general evidence
that iron lines in Seyfert~1 galaxies are indeed broad (Nandra et al. 1997a)
indicates that the iron line emitting region is located close to the
black hole.

An accretion disk can be an effective low-energy reflector if it is
substantially ionized (Matt et al. 1993; \.{Z}ycky et al. 1994).
This may provide a clue to link the hard
X-rays and the soft X-ray excesses.
A hardening of the spectrum above a few keV
(Matt et al. 1993; \.{Z}ycki et al. 1994) {\it and} iron lines
corresponding to He- or H-like states (Matt et al. 1993;
\.{Z}ycki \& Czerny 1994)
are clear signatures of an ionized disk.
There is currently no compelling evidence for ionized
iron lines in the ``broad-line'' Seyfert 1s (Nandra et al. 1997a), but 
there is growing evidence for them in the spectra of Narrow Line
Seyfert Galaxies (NLSy1, Comastri et al. 1998; Turner et al. 1998),
and intermediate luminosity quasars (Nandra et al. 1996).

The Seyfert 1
Galaxy \src\ (${\rm z = 0.038}$) is one of the few objects
where the spectral and temporal behavior from UV and X-ray
observations can be self-consistently explained by an ionized accretion
disk.
The soft X-ray spectrum of \src\ is highly variable,
both in intensity and spectral shape.
The source was discovered during the all sky survey by the Low 
Energy Detectors
on board HEAO-1 and was identified with 
a Seyfert 1 galaxy with an unusually steep
(photon index ${\rm \Gamma = 4.2 \pm 0.6}$) and 
intense [${\rm F_{0.5-4.5 \ keV} =
(3.4 \pm 0.6) \times 10^{-11}}$~erg~cm$^{-2}$~s$^{-1}$]
X-ray spectrum following {\it Einstein} IPC and HRI observations
(Pravdo et al. 1981).
The source was observed by EXOSAT to be in a 100 times
fainter state than observed by HEAO-1 (Piro et al. 1988).
The EXOSAT spectrum was similar to that of other
Seyfert galaxies. Piro et al. (1988) suggested that this
behavior could be explained
in terms of an highly variable soft X-ray excess,
which dominates the emission at high luminosities.
A campaign of simultaneous 
observations using ROSAT and IUE and nearly-simultaneously
with {\it Ginga} (Piro et al. 1997, Paper~I)
as well as a ROSAT observation in 1992,
found the source at an intensity 6 times brighter than
observed by EXOSAT (although still much
fainter than at the time of the HEAO-1 observation). 
These observations confirmed the presence of a variable
soft excess, whose relative strength compared to the
underlying power-law increases with luminosity.
Self-consistent accretion disk models can reproduce
the observed multiwavelength spectra in a low accretion rate
regime (${\rm \dot{\rm M} \equiv L_{disk}/L_{Edd} =}$0.03--0.1), where
the soft X-rays are dominated by reflection from the
mildly ionized regions of the disk. 
Interestingly, such a model left positive residuals in the 0.5--0.8~keV
ROSAT Position Sensitive Proportional Counter 
spectrum, which can be fit by an emission line with 
a centroid energy 
${\rm E = 0.66 \pm 0.10}$~keV and an equivalent 
width ${\rm EW = 50 \pm 30}$~eV. Such a feature is consistent
with K${\rm _{\alpha}}$ fluorescent emission
from O~{\sc vii} or O~{\sc viii} (Paper I).

If this scenario is true and the ionizing flux is not lower than in the state
observed by ROSAT, a continuum component above a few keV
(\.{Z}ycki et al. 1994) and ionized iron fluorescent lines
(Matt et al. 1993; \.{Z}ycki \& Czerny 1994) are
expected in the 1--10~keV spectrum. The
ASCA scientific payload has moderately good energy resolution at the 
energy of the iron lines
(${\rm \sim 2\%}$ at 6~keV) and good sensitivity in the whole
0.5--10~keV band. A search for these features was the motivation 
behind an ASCA observation,
and in this {\it Paper} the relevant results are presented.

\section{Observation and data Reduction}

\src\ was observed by ASCA (Tanaka et al. 1994; Makishima et al. 1996)
between 1995 August 20 12:22~UT and August 21 08:03 UT. The ASCA
scientific payload consists of a pair of CCD cameras (Solid-state Imaging
Spectrometer, SIS), which is sensitive in the 
energy range 0.4--10~keV,
and a pair of gas scintillating proportional counters (Gas Imaging
Spectrometer, GIS), sensitive in the 0.6--10~keV energy range. 
SIS data were
acquired in the 2-CCD FAINT mode and converted to BRIGHT mode for scientific
analysis. GIS data were telemetred in the default PH mode. Standard
selections were applied to the data to avoid Earth occultation
(angle between the pointing direction and the Earth's limb ${\rm \ge
10^{\circ}}$), the bright Earth (angle between
pointing direction and day-night terminator ${\rm \ge 20^{\circ}}$)
and particle (momentum associated with Geomagnetic cut-off rigidity
${\rm \ge 6 \ GeV/c}$) contamination.
Additionally, data obtained within 32~s after any South Atlantic Anomaly
passages were discarded.
SIS grade 0, 2, 3 and 4 data were used.
Scientific products were extracted from circular
regions of 4' and 6' radii for the SIS and GIS, respectively. 
Background
subtraction was performed both with spectra extracted from source
free regions of the same field of view, or from blank sky event files
from the same detector extraction region as the source.
The two methods give results consistent within the
statistical uncertainties. The ``field of view'' and
``blank sky'' backgrounds were chosen for the GIS and SIS spectral
analysis, respectively. The ``field-of-view'' procedure is preferred
for the GIS in order to avoid the known contamination in the blank sky
field caused by the presence of the Seyfert 2 galaxy NGC\thinspace6552
(Fukazawa et al. 1994), 
which strongly reduces the
area available for spectra extraction. The ``blank sky'' method was
instead used for the SIS since significant source contamination
occurs unless a very small extraction area
({\it i.e.}: ${\rm \simlt 2'}$) is chosen. 
The energy range used for spectral analysis are 0.57--9~keV and 0.7--10~keV
for the SIS and GIS, respectively.
These were chosen to exclude low ({\it i.e.} ${\rm \simlt 10\%}$)
effective area regions and SIS calibration problems around the
K-edge of neutral oxygen (${\rm E \simeq 0.54 \ keV}$). The total effective
exposure times after screening were 30, 29, 32.5 and 32.5 ks
for SIS0, SIS1, GIS2 and GIS3, respectively.

Data reduction has been performed with {\sc Ftools 3.6} package;
spectral analysis made use of {\sc Xspec 9.0}. 
Version 4.0 of the GIS publicly available redistribution matrices
has been employed. All energies are quoted in the source rest
frame and uncertainties are given 
at 90\% level of confidence for one interesting
parameter (${\rm \Delta \chi^2 = 2.71}$), unless otherwise specified.

\section{Timing analysis}

\begin{figure}
\begin{center}
\epsfig{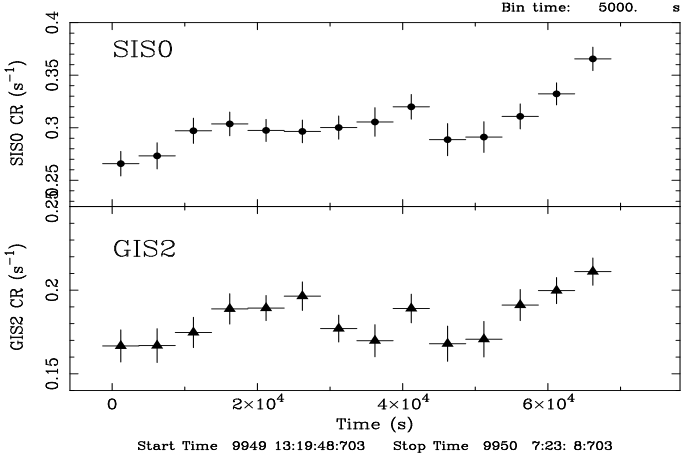}
\end{center}
\caption{Broadband SIS0 and GIS2 light curves. The binning time
is 5000~s}
\label{fig5}
\end{figure}

\begin{figure}
\begin{center}
\hspace{0.75cm}
\epsfig{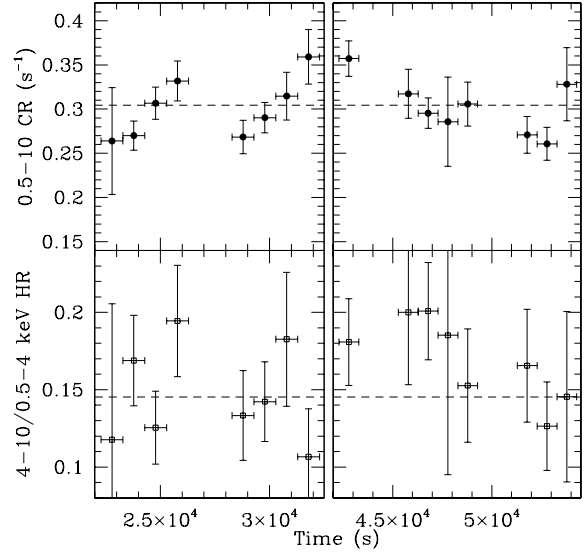}
\end{center}
\caption{Zoom of the SIS0 full band and
4-10/0.5--4~keV HR light curves in the intervals
22000-32500~s ({\it left panels}) and 42500-54500~s ({\it right panels})
after the observation start. The binning time is
1000~s. The average percentage flux increase (decrease) in $10^4$~s
is $90 \pm 40\%$ ($30 \pm 17\%$)}
\label{fig6}
\end{figure}

Figure~\ref{fig5} shows broadband SIS0 and GIS2 light curves.
An overall increase in count rate by a factor ${\rm \simeq 50\%}$ during the
${\rm 7 \times 10^4}$~s observation is evident. We searched for
spectral changes associated with this variation by studying the
hardness ratio (HR) between
the counts in the 0.7--4~keV and 4--9~keV bands (these energy ranges
were chosen
to sample different spectral components, see Sect.~4). The
HR light curve shows no significant variations and the $\chi^2$ for a fit with
a constant straight line is 14.6 for 14 degrees of freedom (dof). The
light curve also displays significant variability episodes on much shorter
time scales. Figure~\ref{fig6} illustrates two such events
as they were observed by
the SIS0 instrument. The implied minimum doubling/halving time is 
1--3${\rm \times 10^4}$~s. The HR seems not to be strongly
correlated with the total flux on short timescales ({\it i.e.}: a few
$\sim 10^3$~s)
either. Better than available
statistics is however needed to confirm this suggestion.

\section{Spectral analysis}

\begin{figure}
\begin{center}
\epsfig{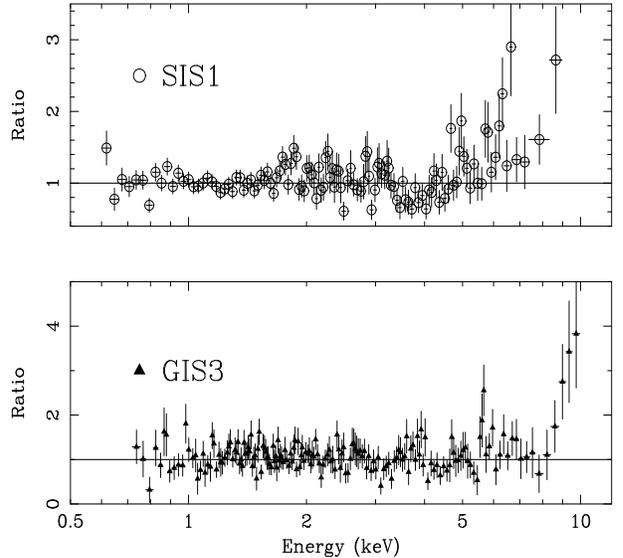}
\end{center}
\vspace{4.0cm}
\caption{Data/model ratio when a simple power-law model is
applied to the spectra of the four detectors simultaneously. Only SIS1
({\it upper panel}) and GIS3 ({\it lower panel}) data are shown.
Energies are plotted in the observer frame. Each data
point has a signal to noise ratio $>3$}
\label{fig1}
\end{figure}

The spectra of the four detectors were fit simultaneously,
after checking that the results of fits using individual spectra are
consistent within the statistical uncertainties.
The data/model ratio is shown in Fig.~\ref{fig1}
when a simple power-law model
absorbed by cold matter is applied.
A constant normalization factor between the SIS and GIS instruments was
left as a free parameter during the initial fits, but it turned out to be
consistent with unity
(${\rm N_{GIS}/N_{SIS} = 0.99^{+0.03}_{-0.02}}$) and it was not included in
subsequent fits. The $\chi^2$ is unacceptable (781 for 659 dof) and is
mainly due to a clear change in the spectral shape at energies
$\approxgt$4~keV. This spectral hardening can be phenomenologically
described using a broken power-law, with spectral indices ${\rm \Gamma_{soft}
= 1.86 \pm 0.07}$ and ${\rm \Gamma_{hard} = 0.9 \pm 0.3}$ and
a break energy ${\rm E_{break} = 4.1^{+0.4}_{-0.6}}$~keV
to give a $\chi^2$ of 718 for 657 dof.
The addition of an absorption edge with a threshold energy
of 7.5~keV, or an emission line with a 
centroid energy of 6.58~keV improves
the quality of the fit marginally (${\rm \Delta \chi^2 = 4.7}$ and 3.0,
respectively). This 
suggests that further spectral complexity is present in the 7--8~keV band.
The 2--10~keV flux is ${\rm 8 \times
10^{-12}}$~erg~cm$^{-2}$~s$^{-1}$, corresponding to a rest frame luminosity
$\sim$${\rm5 \times 10^{43}}$~erg~s$^{-1}$. This is comparable to
the level observed by {\it Ginga} five years earlier (Paper I).

\begin{figure}
\begin{center}
\epsfig{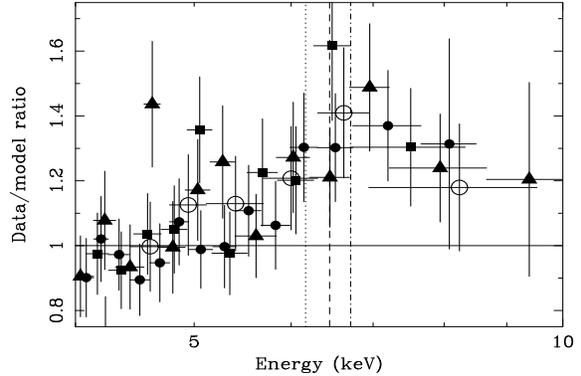}
\end{center}
\caption{Data/model spectral ratio when a power-law and ionized reflection
model is applied to the data of the four ASCA instruments simultaneously.
SIS0 ({\it filled circles}), SIS1 ({\it squares}), GIS2 ({\it triangles}),
GIS3 ({\it open circles}). Each data point has a signal to noise ratio $>$5.
Energies are in the observer frame. The energies of
K$_{\alpha}$ fluorescent emission for neutral ({\it dotted line}),
He-like ({\it dashed line}) and H-like iron ({\it dashed-dotted line})
are indicated}
\label{fig7}
\end{figure}

Spectral hardening above a few keV in Seyfert 1 galaxies
has been interpreted
as the effects of a Compton reflected component arising from
the reprocessing of the nuclear continuum by optically thick matter
around the central black hole.
In the case of \src, this interpretation is further supported 
by the suggestion that
reprocessing from a mildly ionized accretion disk is the most
likely explanation for the UV/soft X-ray emission when the source
was at a comparable luminosity level as in the ASCA observation (Paper I).
We have tested this hypothesis with the
{\sc Xspec} model {\verb!pexriv!} (Magdziarz \& Zdziarski 1995),
which calculates in a self-consistent manner the spectrum produced via
Compton reflection of a power-law primary continuum
by a plane slab.
The model depends on several parameters. The inclination angle ${\rm \theta}$
between the line of sight and the normal to the slab has
been fixed to 30$^{\circ}$ (see a later discussion on the
iron line profile properties for a justification of this choice).
The iron abundance has been assumed cosmic.
The thermal and ionization state of the reflecting material is
parameterized through its temperature ${\rm T}$ and the ionization
parameter ${\rm \xi \equiv L/n r^2}$, 
where ${\rm L}$ is the incident luminosity,
${\rm n}$ the electron number density and ${\rm r}$ the distance between
the primary
continuum source and the reflecting material.
The model is
too complex and the number of free parameters too large
for the available statistics.
If we allow {\it both} ${\rm \xi}$ and ${\rm T}$ to vary freely, they
are not significantly constrained by the data. That is unsurprising, since
varying both parameters affects the detailed shape of the spectrum
below the O~{\sc vii} edge
and above the iron k-edges ({\it i.e.} $\simlt 0.7$~keV and
$\simgt 7$~keV in the observer frame), where the effective
area of the ASCA detectors is low.
The possible presence of an emission line, consistent with
K$_{\alpha}$ fluorescence from Fe~{\sc xxiii}, suggests that
the temperature is as high as ${\rm 4-7 \times 10^5}$~K,
provided that ${\rm \xi}$ is in the range 60--300.
We have therefore fixed ${\rm T}$ at
${\rm 5 \times 10^5}$~K. The corresponding
1$\sigma$ confidence interval on ${\rm \xi}$ is
40--100.
Adding a narrow Gaussian emission line to the continuum
significantly improves the quality of the fit (${\rm \Delta \chi^2 = 14}$).
The centroid energy is ${\rm E = 6.58^{+0.13}_{-0.09}}$~keV,
corresponding to an iron ionization stage ${\rm \ge}$~{\sc xix}.
The profile of the iron line is shown in Fig.~\ref{fig7}.
The $\chi^2$ in the ``ionized'' case is better than in the
case when the disk is assumed to be {\it a priori} ``neutral''
(${\rm \Delta \chi^2 = 6}$ if ${\rm \xi}$ is held fixed to
zero, significant at 98.0\% level).
The improvement is even more significant
when the emission line is included (${\rm \Delta \chi^2 = 14}$, significant
at 99.96\% level).
Moreover, the line best-fit parameters are only slightly affected
by the details of the underlying continuum description, and therefore
the evidence that the line indeed comes from highly ionized species of
iron is rather robust. 
We therefore assume in the following that the continuum
is well described by the combination of a directly observed
power-law and a reflection
spectrum from an ionized disk 
with ${\rm T = 5 \times 10^5}$~K and ${\rm \xi = 80}$.
Table~\ref{tab2} summarizes the best-fit parameters and results
for the reflection model fits.

\begin{table*}
\caption{Best-fit parameters for the ionized disk reflection model.
wa~=~cold absorber, pexriv~=~power-law and reflection from an
ionized disk, ga~=~Gaussian profile, diskline~=~model for a
relativistic Gaussian profile. ${\rm N_{H_{exc}}}$ is the absorption
additional to the galactic value of  $4.2 \times 10^{20}$~cm$^{-2}$}
\begin{tabular}{lccccccc} 
\hline\noalign{\smallskip}
Model & ${\rm N_{H_{exc}}}$ & $\Gamma$ & $R$ & ${\rm E_c}$ 
& $\sigma_1$ & EW & 
$\chi^2/$dof \\ 
& ($10^{20} \ cm^{-2}$) & & & (keV) & (eV) & (eV) \\ 
\noalign {\smallskip}
\hline\noalign {\smallskip}
{\verb!wa*pexrav!} & $< 1.0$ & $1.77\pm0.03$ & $3.7^{+1.0}_{-0.9}$ & ... 
& ... & ... & 732/657 \\
{\verb!wa*(pexrav+ga)!} & $< 1.5$ & $1.78^{+0.04}_{-0.03}$ & 
$3.6^{+1.0}_{-0.9}$ & $6.58^{+0.49}_{-0.10}$ & 0$^{\dag}$ &$150\pm60$ 
& 726/656 \\
{\verb!wa*pexriv!}$^a$ & $< 1.5$ & $1.75^{+0.04}_{-0.03}$ & 
$1.8^{+2.2}_{-0.5}$ & ... & ... & ... & 726/657 \\
{\verb!wa*(pexriv+ga)!}$^a$ & $< 2.0$ & $1.77\pm0.04$ & $1.8^{+2.2}_{-0.5}$ 
& $6.56^{+0.13}_{-0.09}$ & 0$^{\dag}$ &$180^{+80}_{-90}$ & 712/655 \\
{\verb!wa*(pexriv+ga)!}$^a$ & $< 3.7$ & $1.78\pm0.03$ & $1.8\pm0.5$ & 
$6.7\pm0.3$ & 0.43$^{\dag}$ &$420\pm150$ & 705/655 \\
{\verb!wa*(pexriv+ga)!}$^b$ & $< 5.5$ & $1.85\pm0.08$ & $0.4^{+0.6}_{-0.3}$ 
& 6.97$^{\dag}$ & $1.8^{+1.0}_{-0.5}$ & $2500^{+1800}_{-900}$ & 715/656 \\
{\verb!wa*(pexriv+ga+ga)!}$^a$ & $<2.5$ & $1.78^{+0.02}_{-0.04}$ 
& $1.7^{+0.8}_{-0.5}$ & $6.59^{+0.15}_{-0.10}$ & 0$^{\dag}$ 
& $200\pm90$ & 705/654 \\
& & & & $6.01\pm0.15$ & 0$^{\dag}$ & $80\pm50$&  \\
{\verb!wa*(pexriv+diskline)!}$^a$ & $< 0.9$ & $1.82^{+0.08}_{-0.05}$ 
& $1.8^{+0.8}_{-0.5}$ & $6.8^{+0.4}_{-0.3}$ & ... & $590^{+150}_{-180}$ 
& 700/656 \\ 
\noalign {\smallskip}
\hline
\multicolumn{8}{l}{\footnotesize $^{\dag}$held fixed} \\
\multicolumn{8}{l}{\footnotesize$^a$$T \equiv 5 \times 10^5 \ K$ 
fixed, $\xi = 80$} \\
\multicolumn{8}{l}{\footnotesize$^b$$T \equiv 10^7 \ K$, 
$\xi = 10^4$, both fixed} \\
\end{tabular}
\label{tab2}
\end{table*}

\begin{figure}
\begin{center}
\epsfig{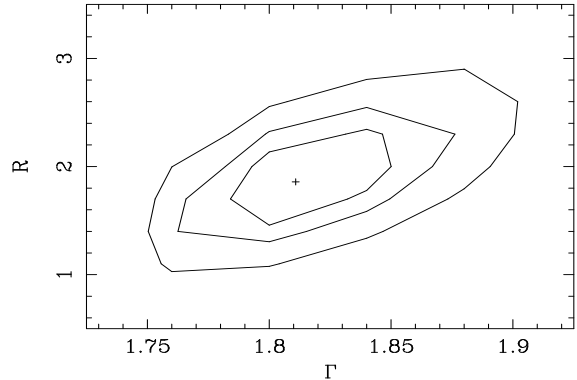}
\end{center}
\caption{$\Gamma$ versus $R$ contour plot. 68\%, 90\% and 99\%
level of confidence iso-$\chi^2$ curves are shown}
\label{fig2}
\end{figure}

Figure~\ref{fig2} shows the contour plot of R 
(the relative normalization between the reflected and primary
continua normalizations) versus $\Gamma$. At the
99\% level of confidence for two interesting parameters, the reflection
is stronger than expected by a plane-parallel infinite slab ({\it i.e.:}
${\rm R = 1}$), whereas ${\rm \Gamma}$ lies in the range
1.74--1.9. The value of ${\rm R}$ is in principle
correlated with the inclination.
However, if ${\theta}$ is fixed to values in the range 0--60$^{\circ}$
(reasonable for a Seyfert 1)
the best-fit nominal values of ${\rm R}$ vary
correspondingly in the range
1.7--2.3, well within the statistical uncertainties.
The absorption cold column density of
${\rm N_H = 5.1^{+2.1}_{-0.9} \times 10^{20}}$~cm$^{-2}$
is consistent with the Galactic
contribution along the line of sight to \src\
($N_H = 4.2 \times 10^{20}$~cm$^{-2}$, Dickey \& Lockman 1990).

We also investigated whether the iron feature is consistent with
multiple components or a broadened structure. Such features are 
expected if the emission region is
located well within the gravitational field of a supermassive black hole.
If the width of the line if left free, then 
there is a significant
improvement in fit quality (${\rm \Delta \chi^2 = 25}$),
with ${\rm E \simeq 7.1}$~keV and ${\rm \sigma \simeq 1.6}$~keV. However, 
such a high centroid energy implies that the iron should be almost 
fully in the
He-like stage. We therefore performed a new fit fixing the
disk parameters to the appropriate values
(${\rm \xi = 10^4}$, ${\rm T = 10^7}$~K). 
The line centroid energy was
fixed at 6.97~keV, corresponding
to fluorescent emission from Fe~{\sc xxvi}.
The $\chi^2$ is comparable to that given by the narrow line
fit ($\chi^2$ =717 for 656~dof). However,
the best-fit width ($\simeq 1.8$~keV)
and the
equivalent width (${\rm EW \simeq 2.5}$~keV) are unplausibly high, 
while typical
values for lines emitted by ionized disks are at most a few hundred
eV (Matt et al. 1993). Unsurprisingly, the amount of
reflection is more than a factor of 3 lower than in the models
where the emission line is assumed to be narrow.
We regard therefore such a solution as unphysical, the line
profile adapting itself easily to the spectrum curvature change due
to the underlying continuum hardening.
If we employ an intrinsic width of 0.43~keV, equal to the average value
measured in the Nandra et al. (1997a) sample, the line turns out to be
slightly more ionized (${\rm E \simeq 6.7}$~keV) and intense (${\rm EW \simeq
420}$~eV) than in the narrow line case.

\begin{figure}
\begin{center}
\epsfig{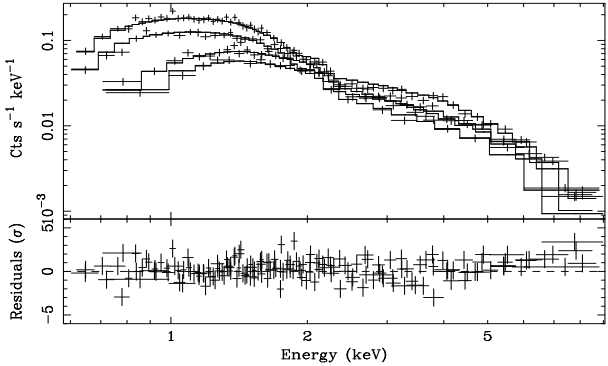}
\end{center}
\caption{Spectra ({\it upper panel}) and residuals in units
of standard deviations ({\it lower panel}) when an ionized disk and
relativistic line model is applied to the ASCA data. All detector
spectra are shown simultaneously. Energies are
in the observer frame. Each data point corresponds to a signal to
noise ratio $>10$}
\label{fig3}
\end{figure}

Alternatively, the addition of a second narrow
Gaussian profile yields a marginally
significant improvement in fit quality (${\rm \Delta \chi^2 = 6}$).
The best-fit parameters of this second component 
are ${\rm E_2 \simeq 6.01 \pm 0.15}$~keV and 
${\rm EW_2 \simeq 80 \pm 50}$~eV. 
It is tempting to associate the second component with
the ``red-horn'' of a gravitationally- and dynamically- distorted Gaussian
profile. In order to test this hypothesis, 
we fit the emission
complex with the {\verb!diskline!} model.
The emissivity law coefficient was fixed at
-2 and the inner and outer radii of the emission region were set
to 10 and 40 Schwarzschild radii, following Paper~I.
We assumed the inclination angles in the {\verb!pexriv!} and
{\verb!diskline!} models to be the same.
The resulting $\chi^2$ of 700 for 654~dof is better than the
double-Gaussian model. The rest energy of the line photons
corresponds to highly ionized iron (${\rm E \simeq 6.8}$~keV)
and the EW is in the range 400--750~eV.
If the inclination angle
is left free to vary, a very marginal improvement of the quality of
the fit is achieved (${\rm \Delta \chi^2 \le 1}$). However, 
we can set a 90\% lower limit on the inclination angle of 22.5$^{\circ}$,
the other line and continuum parameters are only marginally affected
by its precise value.
The spectrum, best-fit model and residuals are shown in Fig.~\ref{fig3}.

We also searched for the low energy emission line detected in
ROSAT data (Paper~I). Adding a further narrow 
Gaussian profile to our data does not 
produce any improvement in $\chi^2$. If we fix the centroid energy at
${\rm E = 0.66 \ keV}$, the 90\% confidence 
upper limit to the  EW is 20~eV.

There are a number of other possible different mechanisms that could
explain the spectral change observed $\approxgt$4~keV.
A hard component leaking through a patchy absorber
could produce a soft excess at energies below the cut-off
corresponding to the absorbing matter density.
However, fits using this model do not
converge, and reduce to the single absorber case. If
a blackbody component is added to the simple absorbed power-law a significant
improvement of the fit is obtained (${\rm \Delta \chi^2 = 31}$), 
but both the
very hard photon index (${\rm \Gamma \simeq 1.51}$) and the rather 
high temperature
(${\rm kT \simeq 280 \ eV}$) are hard to reconcile with our present 
understanding
of physical processes in the nuclear region of AGN and with the available
observational results on Seyfert~1 X-ray spectra. 

\section{Discussion}

\src\ displays a very complex and extreme soft X-ray behavior,
with intensity variations by one order of magnitude on
a timescale of years (Pravdo et al. 1981; Piro et al. 1988).
In Paper I, Piro et al. (1997) proposed that an ionized disk plays an
important role in determining the soft X-ray spectrum of this source.
A low ionization state disk would produce the very dim states observed,
for instance, by EXOSAT. A mildly ionized disk (${\rm \xi \sim 100}$)
in a low accretion rate regime (${\rm \dot{\rm M} < 0.1}$) could explain the
intermediate states, as observed in the multiwavelength IUE-ROSAT-{\it Ginga}
campaign (Paper I). In that observation, an emission line was discovered
with a centroid energy consistent with K$_{\alpha}$ fluorescence from
O~{\sc vii} or O~{\sc viii}. For higher accretion rates, the disk would
become even more highly ionized and one could expect an ionized
iron line to appear in the X-ray spectrum.

This prediction is nicely fulfilled by the detection of
an iron emission line in the
X-ray spectrum of \src\ reported here. The centroid energy
in the source rest frame is 6.6--6.8~keV and is
formally inconsistent with emission
from neutral or mildly ionized iron. 
The implied ionization stages correspond
to Fe~$>$~{\sc xix}.
It is the first time that a ``warm'' iron line has been detected in a
``broad-line'' Seyfert 1.
Such a high ionization state can be achieved in
a thermally stable disk, provided that the temperature is $>$$4 \times
10^5$~K and the ionization parameter ${\rm \xi}$ lies between 60 and
500. In the above analysis, the lack of statistics and the limited ASCA
bandpass prevent us leaving all the spectral parameters which
characterize the physical state of the disk free.
However, if we assume a temperature of ${\rm 5 \times 10^5}$~K,
${\rm \xi}$ turns out self-consistently to be $\simeq 80$. The line
profile and continuum shape allow us to set a 90\% lower limit
of 22.5$^{\circ}$ on the inclination
of the disk normal to the line-of-sight.

The observed EW of the line 
ranges between 300 and 600~eV, depending on the
line model used. The observed combination of line
centroid energy and EW
is consistent with the ionized disk scenario provided that
${\rm \dot{M}}$ is the range 0.2--0.4.
No emission line from ionized stages
of oxygen is detected in the ASCA spectra, the upper limit to a
0.66~keV narrow line being inconsistent at the 1$\sigma$ level with the
positive detection using ROSAT reported in Paper I.
This is once again consistent with the ionized disk scenario.
Under the proposed combination of ionization parameter and temperature,
the oxygen atoms are expected to be fully stripped (\.{Z}ycki et al. 1994)
and any relevant line would therefore disappear.

X-ray flux variations with a doubling time of a few thousand 
seconds have been discovered for the first time in \src.
Usual light crossing arguments constraint the mass of the central
accretor to be less than a few~${\rm  \times 10^7}$~${\rm M_{\odot}}$. This is
in broad agreement with the spectral analysis results:
${\rm \dot{M} = 0.3}$ and ${\rm L_X \sim 5 \times 10^{43}}$~erg~s$^{-1}$
implying a central accretor mass of ${\rm \sim 3 \times 10^6 \ M_{\odot}}$.

Coherently with with the iron line properties, the intermediate X-ray continuum
of \src\ is best modeled with a primary
power-law, with photon index, ${\Gamma}$, in the range 1.74--1.9,
and a Compton reflection component from highly ionized matter.
The relative normalization between the reflected and primary continua
is higher than expected from a plane-parallel slab, subtending a
${\rm 2 \pi}$ angle from the central source. Several effects could account
for this. The reflecting matter could be arranged in a warped
geometry, or the primary radiation could be anisotropic
and therefore the disk could see more flux than directed along our
line of sight. In principle, the same effect could be produced
by a lag in the response of the reflector
to any changes in the ionizing flux,
longer than the short-term variability \src\ time scales.
Since the average ${\rm R}$ inferred from the
ASCA data is $>$1, while the primary continuum {\it increases} by 40\%,
one has to assume that the reflector is responding to a brighter state
that occurred before the ASCA observation started. No spectral hardening
occurred during the ${\rm 6.5 \times 10^4}$~s duration of the ASCA
observation.
The nuclear continuum should then be reproducing a very similar pattern
of variability during the ASCA observation and during the previous
brighter state. If this is indeed the case,
the bulk of the reflecting matter should then be located
$\simgt$$10^{15}$~cm from the nucleus, or ${\rm \simgt 10^3 R_S}$
for a black hole with a mass of ${\rm 3 \times 10^6 M_{\odot}}$. 
However, this conclusion
is almost at odds with the evidence, again coming from the ASCA data,
that the line is broad, as expected if it is
produced in the immediate vicinity of a supermassive black hole
(Matt et al. 1992). The indication that the Compton
reflected component probably follows quasi-simultaneously
the short-term ({\it i.e.}: $\sim$ a few $10^3$s) variations of the
primary flux (see Fig.~\ref{fig6}) further supports the idea that the
bulk of the ionized reflecting matter is located within $\sim 10^2$
gravitational radii from the nucleus.

The ASCA observation has provided the first simultaneous measure of the
\src\ spectral properties in the soft
and intermediate X-ray energy bands. It 
impossible to directly compare the secular changes in the properties of
the accretion flow/matter inferred from the oxygen and iron line
properties with the contemporary luminosity of the ionizing continuum.
If we compare the spectral shape measured by ROSAT and ASCA
in the 0.5--2.5~keV energy band of overlap, the former turns out to be
much steeper 
(${\rm \Gamma_{ROSAT} \simeq 2.7}$, ${\rm \Gamma_{ASCA} \simeq 1.8}$).
This effect is significantly higher than the well-known systematic
higher steepness of ROSAT spectra in comparison with ASCA ones
(Fiore et al. 1994; Iwasawa et al. 1998).
This again is in good agreement with the expectation of the
ionized reflection scenario (Matt et al. 1993), which
predicts a flatter and more featureless spectrum in that band,
for increasing values of ${\rm \dot{M}}$.
Moreover, the higher the accretion rate, the less prominent
any soft X-ray excess should be,
which is in good agreement with the lack of a strong
soft excess in the ASCA observation (see Fig.~\ref{fig1}).
The soft X-ray state observed by ROSAT
is more luminous than ASCA's (${\rm F^{ROSAT}_{0.5-2.5 \ keV} \simeq
6.8 \times 10^{-12}}$~erg~cm$^{-2}$~s$^{-1}$; 
${\rm F^{ASCA}_{0.5-2.5 \ keV} \simeq
4.3 \times 10^{-12}}$~erg~cm$^{-2}$~s$^{-1}$).
This apparent contradiction with the model expectations
can be easily resolved
if one assumes that the intermediate X-ray ionizing continuum was
at the epoch of ROSAT observation about a factor
of 3 higher than observed by ASCA. The dynamical range of the
historical 2--10~keV flux is about 6 (see Paper I) and therefore
of the required order of magnitude.

\src\ displayed an unusual soft state during part of the 1980s.
The source detected by HEAO-1 exhibited a power-law with a photon index
${\rm \Gamma \simeq 4.2}$.
The 0.5--4.5~keV
flux was about a factor of five larger than observed in the same band by ASCA
(${\rm \sim 6.7 \times 10^{-12}}$~erg~cm$^{-2}$~s$^{-1}$).
In Paper I it was speculated that a highly ionized disk
with an accretion rate close to the Eddington limit could explain
the extremely high soft X-ray state observed by HEAO-1, due to the
combined effects of bremsstrahlung emission from hot electrons produced
by the ionizing continuum, of the reflected continuum and of
the direct emission from the accretion disk. 
This seems not to be the case in the ASCA
observation. The power-law photon index in the 0.7--4~keV band is
${\rm \Gamma \simeq 1.9}$, in good agreement with that typically observed in
Seyfert 1 galaxies (Nandra et al. 1997a) and no further soft excess
is required in addition to the reflected power-law model.
However, a steeper and brighter soft X-ray
emission can be obtained in this framework with an appropriate
combination of slightly higher ${\rm \xi}$ ($> 300$, \.{Z}ycki et al. 1994:
Paper I) and steeper intrinsic continuum.

An alternative explanation to obtain short ``bursts'' of accretion
invokes instabilities in the disk. Such models
have been proposed to explain the optical/UV variability in AGN
(Siemiginowska \& Czerny 1989) and the giant X-ray outburst
in some NLSy1 (Grupe et al. 1995a,b). An instability at radial
distance ${\rm r}$ can determine a switch from a low to a high viscosity
state and induce a rapid accretion phase, which decays over the
viscous timescale ${\rm t_{visc} \sim 1.7 \times 10^9 \ \alpha^{-4/5}
\dot{M}_{16}^{-3/10} \ M_1^{1/4} r^{5/4}_{13}}$~s 
(Frank et al. 1975),
where $\alpha$ is the viscous 
parameter, ${\rm \dot{M}_{16}}$ is the accretion rate in units of
${\rm 10^{16}}$~g~s$^{-1}$, 
${\rm M_1 = M/M_{\odot}}$ and ${\rm r_{13} = r/(10^{13}}$~cm).
If we assume, following Matt et al. (1993), ${\rm \alpha = 0.1}$,
${\rm \dot{M} = 0.2}$ and ${\rm M_1 \sim 3 \times 10^6}$, it follows that 
${\rm t_{visc} \sim
2 r^{5/4}_{13}}$ years, what makes such a mechanism viable to account for the
observed historical soft X-ray behavior of \src. The transient nature
of near-Eddington accretion events has already been proposed as a possible
explanation for the high dispersion in the $\alpha_X$ distribution
in soft X-ray selected AGN (Grupe et al. 1998).

In none of the objects of the Nandra et al. (1997a) sample of Seyfert 1s
observed by ASCA is the centroid of the iron line inconsistent with
emission from neutral or mildly ionized iron. However, an ionized iron
emission line was detected in the NLSy1 TonS180
(Comastri et al. 1998; Turner et al. 1998).
NLSy1 are characterized by steep soft
X-ray spectra (B\"oller et al. 1996) and extreme soft X-ray
variability (B\"oller et al. 1993, 1997; Grupe et al. 1995a,b;
Forster \& Halpern 1996).
The former property bears a resemblance with the X-ray
spectra of Galactic black hole candidates in their high and
soft state, suggesting that soft X-rays in NLSy1 galaxies are
produced by intrinsic emission from a quasi-Eddington
accretion disk (Pounds et al. 1995). If this is indeed the case,
the surface of the disk
should be strongly ionized (Matt et al. 1993) and iron K$_{\alpha}$
fluorescent line from highly ionized iron species are expected.
This scenario also explains naturally the clear anti-correlation
between the soft X-ray spectral index and the width of the
H$_{\beta}$ line (B\"oller et al. 1996). This relation
would be the mirror of the
${\rm L/L_{Edd} \propto v^{-2}}$ scale law, which arises
if the broad lines are produced in a virialized gas and the
Broad Line Region size is determined uniquely by the nuclear luminosity.
In this picture, the accretion rate is the only physical
parameter, which determines both the X-ray and the optical properties.
Unfortunately,
no optical observation simultaneous with
the extreme soft state observed in \src\ by HEAO-1 exist.
A spectrum taken two years later showed relatively broad
H$_{\beta}$ (${\rm FWHM \sim 4000}$~km~s$^{-1}$; Pravdo et al. 1981).
\src\ could undergo transitions between ``narrow''
and ``broad'' Seyfert 1 states, driven by changes of its accretion
rate from few tenths to quasi-Eddington values. Future
simultaneous optical and X-ray monitoring of \src\ are needed
to test this intriguing hypothesis.

The most extreme example of soft X-ray variability in NLSy1s
is IRAS{\thinspace}13224$-$3809, which exhibits variations up to a factor
$\sim$57 in 2 days (B\"oller et al. 1997).
The most convincing explanation for such a
phenomenology is Doppler boosting, induced by relativistic
motions in the inner regions of an accretion disk
A boost factor
of 5 for ${\rm \Gamma \simeq 4}$ can be achieved if the system is seen at
an inclination ${\rm \simgt 40^{\circ}}$ (Guilbert et al. 1983).
This mechanism is in principle also viable to explain the
exceptional X-ray soft states of 1E1615+061. However,
the much steeper spectrum observed by HEAO-1 remains
to be explained, and the interplay between spectral
variability and flux amplification would make it difficult to disentangle
the two causes.

It is worthwhile to note that the
2-10~keV luminosity is not unusual compared with other ``broad''
Seyfert 1s (${\rm \log(L_X) \sim 43.70}$) and therefore the argument that
ionized disks occur preferentially in high-luminosity systems
(Nandra et al. 1996; Nandra et al. 1997b) cannot
be applied to \src.

Other explanations for the huge soft X-ray variability can be
discarded at the light of the ASCA outcomes. The lack of any absorption
feature from ionized oxygen (which confirms the first evidence emerging from
the ROSAT spectrum, Paper I) excludes a significant contribution from
an ionized absorber to the \src\ line of sight opacity, whose
changes could mimic the appearance and variability of a soft excess.
Alternatively, one could argue that the steeper spectrum
observed in the 80s was due
to a change of the nuclear primary continuum. This hypothesis has
been recently invoked to explain the extreme soft X-ray variability
in the otherwise standard Seyfert 1 1H{\thinspace}0419$-$577 
(Guainazzi et al. 1998).
However, a ${\rm \Gamma = 4}$ spectrum would require very extreme conditions
in the Comptonizing plasma, either in terms of very low electron
density temperature or very high optical thickness (Hua \& Titarchuk 1995).
Such conditions cannot be achieved in the standard two 
phase corona-disk model (Haardt \& Maraschi 1993; Haardt et al. 1997).

The new X-ray results on \src\ which provide the first
direct evidence of reprocessing from highly ionized matter in the
immediate proximity of a supermassive black hole in the 
nuclear regions of Seyfert 1 galaxies, 
confirm
the general importance of reprocessing phenomena in the AGN environment.
Future attempts to follow the multiwavelength history of this,
and similar, objects will certainly very fruitful and allow
better constraints to be set on the physics and geometry of the matter located
in the radiation field of these monsters.
 
\begin{acknowledgements}

MG acknowledges an ESA Research Fellowship. The careful reading by two
referees allowed us to improve the organization of the
paper and the comprehensiveness of the discussion.

\end{acknowledgements}


\begin{thebibliography}{}

\bibitem{} Betchold J., Elvis M., Fiore F., et al., 1994, AJ 108, 374 

\bibitem{} B\"oller T., Tr\"umper J., Molendi S., et al., 1993, A\&A 279, 53 

\bibitem{} B\"oller T., Brandt W.N., Fabian A.C., 1997, MNRAS 289, 393 

\bibitem{} B\"oller T., Brandt W.N., Fink H.H., 1996, A\&A 305, 53 

\bibitem{} Clavel J., Reichert G.A., Alloin D., et al., 1991, ApJ 366, 64 

\bibitem{} Comastri A., Fiore F., Guainazzi M., et al., 1998, A\&A 333, 31

\bibitem{} Czerny B., Elvis M., 1987, ApJ 321, 305 

\bibitem{} Dickey J.M., Lockman F.J., 1990, ARA\&A 28, 215

\bibitem{} Fiore F., Elvis M., McDowell J.C., et al., 1994, ApJ 431, 515

\bibitem{} Forster K., Halpern J.P., 1996, ApJ 468, 565

\bibitem{} Frank J., King A.R., Raine D.J., 1975, ``Accretion power in 
Astrophysics'', Cambridge University Press, p.101

\bibitem{} Fukazawa Y., Makishima K., Ebisawa K., et al., 1994, PASJ 46, L141 

\bibitem{} George I., Fabian A.C., 1991, MNRAS 249, 352

\bibitem{} Grupe D., Beuermann K., Mannheim K., et al., 1995a, A\&A 299, L5 

\bibitem{} Grupe D., Beuermann K., Thomas H.-C., et al., 1995b, A\&A 300, L21

\bibitem{} Grupe D., Beuermann K., Thomas H.-C., Mannheim K., Fink H.H., 
1998, A\&A 330, 25

\bibitem{} Guainazzi M., Comastri A., Stripe G., et al., 1998, A\&A, in press
(available at astroph/9808009)

\bibitem{} Guilbert P.W., Fabian A.C., Rees M.J., 1983, MNRAS 205, 593

\bibitem{} Haardt F., Maraschi L., 1993, ApJ 413, 507

\bibitem{} Haardt F., Maraschi L., Ghisellini G., 1997, ApJ 476, 620

\bibitem{} Hua X.-M., Titarchuk L., 1995, ApJ 449, 188

\bibitem{} Iwasawa K., Brandt W.N., Fabian A.C., 1998, MNRAS 293, 251

\bibitem{} Kuhn O., Betchold J., Cutri R., Elvis M., Rieke M., 1995, ApJ 438, 643 

\bibitem{} Laor A., 1990, MNRAS 246, 396 

\bibitem{} Laor A., Fiore F., Elvis M., et al., 1997, ApJ 477, 93 

\bibitem{} Magdziarz P., Zdziarski A.A., 1995, MNRAS 273, 837 

\bibitem{} Makishima K., Tashiro M., Ebisawa K., et al., 1996, PASJ 48, 171

\bibitem{} Matt G., Perola G.C., Piro L., Stella L., 1992, A\&A 257, 63

\bibitem{} Matt M., Fabian A.C., Ross R.R., 1993, MNRAS 264, 839

\bibitem{} Nandra K., Pounds K.A., 1994, MNRAS 268, 405 

\bibitem{} Nandra K., George I.M., Turner J.T., Fukazawa Y., 1996, ApJ 
464, 165 

\bibitem{} Nandra K., George I.M., Mushotzky R.F., Turner T.J., Yaqoob T., 
1997a, ApJ 477, 602

\bibitem{} Nandra K., George I.M., Mushotzky R.F., Turner T.J., Yaqoob T., 
1997b, ApJ 488, L91 

\bibitem{} Peterson B.M., Bolonek T.S., Barker E.S., et al., 1991, ApJ 
368, 119 

\bibitem{} Piro L., Massaro E., Perola G.C., Molteni D., 1988, ApJ 325, L25

\bibitem{} Piro L., Yamauchi M., Matsuoka M., 1990, ApJ 360, L35 

\bibitem{} Piro L., Ba\l uci\'nska-Church M., Fink H.H., et al., 1997, 
A\&A 319, 74 (Paper~I)

\bibitem{} Pounds K., Nandra K., Stewart G.C., George I.M., Fabian A.C., 
1990, Nat 344, 132

\bibitem{} Pounds K., Done C., Osborne J., 1995, MNRAS 277, L5

\bibitem{} Pravdo S.H., Nugent J.J., Nousek J.A., et al., 1981, ApJ 251, 501

\bibitem{} Puchnarewicz E.M.P., Mason K.O., Romero-Colmenero E., et al., 
1996, MNRAS 281, 1243 

\bibitem{} Siemiginowska A., Czerny B., 1989, MNRAS 239, 289

\bibitem{} Tanaka Y, Inoue H., Holt S.S., 1994, PASJ 46, L37 

\bibitem{} Tanaka Y., Nandra K., Fabian A.C., et al., 1995, Nat 375, 659

\bibitem{} Turner T.J., George I.M., Nandra K., 1998, ApJ in press
(available at astroph/9806323)

\bibitem{} Ulrich M.H., Molendi S., 1995, A\&A 293, 641 

\bibitem{} Walter R., Fink H.H., 1993, A\&A 274, 105 

\bibitem{} \.{Z}ycki P.T., Czerny B., 1994, MNRAS 266, 653 

\bibitem{} \.{Z}ycki P.T., Krolik J.H., Zdziarski A.A., Kallman T.R., 
1994, ApJ 437, 597 
 
\end{thebibliography}
\end{document}